\documentclass[fleqn,usenatbib]{mnras}
\usepackage[dvipsnames]{xcolor}

\usepackage{newtxtext,newtxmath}
\usepackage[T1]{fontenc}

\DeclareRobustCommand{\VAN}[3]{#2}
\let\VANthebibliography\thebibliography
\def\thebibliography{\DeclareRobustCommand{\VAN}[3]{##3}\VANthebibliography}

\usepackage{graphicx}	% Including figure files
\usepackage{amsmath,amsfonts}	% Advanced maths commands
\usepackage{color}
\usepackage{ulem}

\newcommand{\modotyr}[1]{M_{\odot}{\rm yr^{-1}}}
\title[Accretion onto wandering IMBHs]{Three-dimensional radiation hydrodynamics simulations of wandering intermediate-mass black holes considering the anisotropic radiation and dust sublimation}

\author[Ogata et al.]{
Erika Ogata,$^{1}$\thanks{E-mail: ogata@ccs.tsukuba.ac.jp}
Ken Ohsuga,$^{1}$
Hajime Fukushima$^{1}$
and Hidenobu Yajima$^{1}$
\\
$^{1}$Center for Computational Science, University of Tsukuba, Ten-nodai, 1-1-1 Tsukuba,  Ibaraki 305-8577, Japan\\
}

\date{Accepted XXX. Received YYY; in original form ZZZ}

\pubyear{2024}

\begin{document}
\label{firstpage}
\pagerange{\pageref{firstpage}--\pageref{lastpage}}
\maketitle

\begin{abstract}
By performing three-dimensional radiation hydrodynamics simulations, we study Bondi-Hoyle-Lyttleton accretion onto intermediate-mass black holes (BHs) wandering in the dusty gas. 
Here, we take into account the anisotropic radiation feedback and the sublimation of dust grains. 
Our simulations show that when the relative velocity between the BH and the gas is small ($\sim 20\rm km~s^{-1}$) and gas density is $\sim 10^4 \rm cm^{-3}$, the gas mainly accretes from near the equatorial plane of the accretion disk at a time-averaged rate of 0.6\% of the Bondi-Hoyle-Lyttleton rate. 
An ionized region like two spheres glued together at the equatorial plane is formed, and the dense shock shell appears near the ionization front. 
The BH is accelerated at $\sim 10^{-8} \rm cm~s^{-2}$ due to the gravity of the shell. 
For denser gas ($\sim 10^6 \rm cm^{-3}$), the time-averaged accretion rate is also 0.6\% of the Bondi-Hoyle-Lyttleton rate.
However, the BH is decelerated at $\sim10^{-7} \rm cm~s^{-2}$ due to gravity of the dense downstream gas although the dense shock shell appears upstream.
Our simulations imply that intermediate-mass BHs in the early universe keep floating at $\gtrsim {\rm several}\, 10\,\rm km~s^{-1}$ without increasing mass in interstellar gas with density of $\sim 10^4 \rm cm^{-3}$, and slow down and grow into supermassive BHs in galaxies with the density of $\sim 10^6 \rm cm^{-3}$.
\end{abstract}


%%%%%%%%%%%%%%%%%%%%%%%%%%%%%%%%%%%%%%%%%%%%%%%%%%

%%%%%%%%%%%%%%%%% BODY OF PAPER %%%%%%%%%%%%%%%%%%

\section{Introduction}
\label{S:Introduction}

Bondi-Hoyle-Lyttleton accretion is the accretion phenomenon that occurs under the condition where there is relative motion between the central object and the surrounding gas, 
in which the gas is attracted by gravity and accretes to the central object
\citep[for recent reviews, see ][]{Edgar2004}.
It occurs
when the gas accretes to neutron stars and stellar-mass black holes (BHs) which are kicked by supernova explosions \citep{Hobbs2005,Wong2010,Janka2012,Wongwathanarat2013,Tauris2017,Kapil2023}, 
to stars and BHs in globular clusters \citep{Portegies2002,Gurkan2004,Kaaz2019,Shi2021,Gonzalez2021}, to compact objects in X-ray binaries via stellar wind accretion, and to intermediate-mass BHs (IMBHs) floating in the remnant galactic disk \citep{Di2005,Di2012,Di2017,Dubois2012,Sijacki2015,Latif2018}.

Bondi-Hoyle-Lyttleton accretion flow was investigated through
numerous hydrodynamics simulations from the 1970s to the late 1990s \citep{Hunt1971,Shima1985,Fryxell1988,Ho1989,Matsuda1991,Ruffert1994,Ruffert1996}, which showed that numerically obtained the accretion rates are consistent with the analytical estimates of \cite{Hoyle1939} and \cite{Bondi1944}.
Subsequently, numerical simulations considering the radiation from the central object were performed \citep{Blondin1990,Taam1991,Milosavljevic2008,Park2012,Park2017,Sugimura2020,Toyouchi2020}, showing that the accretion rates are significantly reduced by radiation feedback e.g., the radiation force and ionization heating.
In the case in which the central object is compact object with an accretion disk, however, the accretion rates can not decrease significantly because the gas accretes through the region near the disk plane, where the radiation effect is relatively weak.
Although Bondi-Hoyle-Lyttleton accretion onto the compact objects under the anisotropic radiation field produced by the accretion disk has been investigated in previous studies \citep{Fukue1999,Hanamoto2001,Ogata2021}, the effect of fluid dynamics such as gas pressure gradient force has not been taken into account.
Therefore, radiation hydrodynamics simulation of Bondi-Hoyle-Lyttleton accretion including an anisotropic radiation field is required.

The increase in the absorption coefficient due to dust also decreases the accretion rate 
because the radiation force becomes more effective.
In fact, \cite{Yajima2017} and \cite{Toyouchi2020} showed that accretion onto the central BHs is suppressed by the strong radiation force acting on dusty gas through radiation hydrodynamics simulations.
\cite{Toyouchi2020} revealed that the accretion rate decreases with increasing metallicity (dust amount) 
and is comparable to the Eddington accretion rate for dusty gas.
In these previous studies, 
dust sublimation was not considered, and isotropic radiation fields were adopted.
More realistically, dust sublimation should also be taken into account, because the radiative force is reduced in dust sublimation regions.

In this study, 
we perform three-dimensional radiation hydrodynamics simulations to investigate the effects of radiation anisotropy and dust sublimation on the Bondi-Hoyle-Lyttleton accretion mechanism.
We focus on the growth process of supermassive BHs observed in the early universe and select IMBHs floating in dusty gas with metalicity $Z=0.1Z_{\odot}$ as our target objects.
The rest of this paper is organized as follows.
We describe the method of our three-dimensional radiation hydrodynamics simulations in Section \ref{S:Simulation Methods}.
The simulation results are given in Section \ref{S:Results},
and we discuss the evolution of IMBHs in Section \ref{S:Discussions}.
Finally, we provide a summary and conclusion in Section \ref{S:Summary and Conclusions}.

\section{Simulation Methods}
\label{S:Simulation Methods}
In this study, we perform three-dimensional radiation hydrodynamics simulations with {\sc SFUMATO-M1} \citep{Fukushima2021}.
This code is based on {\sc SFUMATO-RT} \citep{2020ApJ...892L..14S}, which is the modified version of {\sc SFUMATO} \citep{Matsumoto2007,Matsumoto2015}.
In {\sc SFUMATO-M1}, the module with the M1-closure technique is adapted instead of the adoptive ray-tracing solver developed in \citet{2020ApJ...892L..14S}.
We further include the anisotropic radiation field and dust sublimation for the current purpose.

\subsection{Basic Setup}
\label{S:Basic Setup}

We perform the three-dimensional radiation hydrodynamics simulations with the Cartesian coordinate and simulate the gas flow relative to the BH.
The gas has constant number density $n_{\infty}$, positive $x$-direction velocity $v_{\infty}$, and temperature $T_{\infty}=180~{\rm K}$ at the upstream boundary.
We set the size of a calculation box as $R_{\rm out} = 12.6~{\rm pc}$, which is much larger than Bondi-Hoyle-Lyttleton radius and the size of ionized region around a BH in our simulations.

We adopt the sink method to mask the accretion disk and set the sink radius as $R_{\rm sink}=2.7\times 10^{-3}~\rm pc$, which is small enough to resolve dust sublimation region.
We use the nested grid method to resolve the gas structure inside ionized region and dust sublimation region efficiently.
In our simulations, the maximum refinement level is $l_{\rm max}=10$, and the minimum cell size is $\Delta_{10}=3.85\times 10^{-4}~\rm pc$.
The BH is fixed at the origin and grows due to mass accretion.
However, the increase in the BH mass is negligibly small in the elapsed time of our simulations.
Also, we neglect acceleration of the BH, self-gravity of gas, and magnetic field for simplicity.

In the model with the anisotropic radiation, we perform the convergence check tests with the higher refinement level $l_{\rm max} = 11$.
As a result, we confirm good convergence in terms of the flow structure and the accretion rate.
We also perform the simulations with the same sink size and resolution of \citet{Toyouchi2020}, where they investigated Bondi-Hoyle-Lyttleton accretion of the dusty gas around luminous object by three-dimensional radiation hydrodynamics simulations.
We confirm that our simulation results are consistent with their study.

The M1-closure method employed in this study is known to be more diffusive than the ray-tracing method \citep[e.g.][]{Asahina2020}. 
By the test calculation of an ionized bubble produced by the anisotropic radiation, it is found that the distance from the radiation source to the ionization front using the M1-method is 1.2-1.5 times larger (smaller) than the analytical estimation around the disk equatorial plane (rotation axis).
This discrepancy may be due not only to differences in the calculation methods for radiative transfer but also to differences in the treatment of photons emitted by recombination. 
Our method also takes into account the recombination photons, but the Case-B recombination is assumed in the analytical solution. 
The expansion of the ionized region due to recombination photons is more pronounced near the equatorial plane because many recombination photons enter from the ionized regions sandwiching the equatorial plane. 
Simulations using more accurate radiation transfer methods are left as an important future work \citep[e.g.][]{Sugimura2020}.

%=========================TABLE================================
\begin{table*}
	\centering
	\caption{
	Model parameters with the results of the time-averaged accretion rate and the luminosity.
	}
	\begin{tabular}{p{24mm}ccccccr} % four columns, alignment for each
		\hline
		Model & $n_{\infty}(\rm cm^{-3}$) & $v_{\infty}(\rm km~s^{-1})$ & radiation field & $\overline{{\dot{M}}}~(M_{\odot}\rm yr^{-1})$ & $\overline{L}~(L_{\odot})$ &
		$\overline{{\dot{M}}}/\dot{M}_{\rm BHL}$ \\
		\hline
		isoN4V20 & $10^4$ & $20$ & isotropic & $4.3\times 10^{-6}$ & $6.3\times 10^{6}$ & $6.2\times 10^{-3}$\\
		edgeN4V20 & $10^4$ & $20$ & edge-on & $4.4\times 10^{-6}$ & $6.5\times 10^{6}$ & $6.4\times 10^{-3}$\\
		edgeN6V20 & $10^6$ & $20$ & edge-on & $4.2\times 10^{-4}$ & $4.6\times 10^{8}$ & $6.1\times 10^{-3}$\\
		edgeN4V100 & $10^4$ & $100$ & edge-on & $4.4\times 10^{-6}$ & $6.5\times 10^{6}$ & $5.6\times 10^{-1}$\\
		poleN4V20 & $10^4$ & $20$ & pole-on & $4.2\times 10^{-6}$ & $6.2\times10^{6}$ & $6.1\times 10^{-3}$\\
		\hline
	\end{tabular}
	\label{tab:model}
\end{table*}
%=========================TABLE================================

\subsection{Basic Equations}
\label{S:Basic Equations}
We solve the following governing equations:
the equation of continuity,
\begin{equation}
    \frac{\partial\rho}{\partial t} + \nabla \cdot (\rho \boldsymbol{v}) 
    = 0,
\end{equation}
the equation of motion,
\begin{equation}
    \frac{\partial(\rho \boldsymbol{v})}{\partial t} 
    + \nabla \cdot (\rho \boldsymbol{v} \otimes \boldsymbol{v}) 
    +\nabla P
    = \rho(\boldsymbol{g}+\boldsymbol{f}),
\end{equation}
and the equation of energy,
\begin{equation}
    \frac{\partial(\rho E)}{\partial t} 
    + \nabla \cdot [(\rho E + P)\boldsymbol{v}]
    = \rho(\boldsymbol{g}+\boldsymbol{f})\cdot \boldsymbol{v} + \Gamma -\Lambda,
\end{equation}
where $E$ is total energy defined as
\begin{equation}
    E
    =
    \frac{|\boldsymbol{v}|^2}{2} + (\gamma-1)^{-1}\frac{P}{\rho}.
\end{equation}
Here,
$\rho$, $\boldsymbol{v}$, and $P$ are the gas density, gas velocity, and gas pressure, respectively, 
and $\boldsymbol{g}$ and $\boldsymbol{f}$ are the gravity and radiation force,
and $\Gamma$ and $\Lambda$ are the specific heating and cooling rates. 
We estimate the adiabatic exponent $\gamma$ as in \cite{Omukai1998}.
In this simulations, we also solve the transfer of far-ultraviolet (FUV; $11.2~{\rm eV}$ - $13.6~{\rm eV}$) and extreme-ultraviolet (EUV; $13.6~{\rm eV}$-$1~{\rm keV}$) photons emitted from the sink using the M1-closure approximation method. We also compute the transfer of infrared (IR) photons mainly emitted from dust grains as thermal emission. Further details on our radiative transfer method are described in \citet{Fukushima2021}.

We calculate the non-equilibrium chemical reactions for the eleven species of 
$\rm H, H_2, H^-, H^+, H_2^+, e, CO, C_{II}, O_I, O_{II}$, and $\rm O_{III}$.
The number density of the $i$-th species is calculated as 
\begin{equation}
    \frac{\partial (y_in_{\rm H})}{\partial t}
    +
    \nabla \cdot (y_i n_{\rm H} \boldsymbol{v})
    =
    n_{\rm H}R_i,
\end{equation}
where $y_i=n_i/n_{\rm H}$ is the number ratio of $i$-th species to hydrogen nuclei, 
and $R_i$ is the corresponding chemical reaction rate.
We also consider the dust grains in the gas, 
assuming the dust-to-gas mass ratio of $0.01 \times  Z/Z_{\odot} =10^{-3}$. 
The temperature of dust grains is estimated from the energy balance 
between the absorption/emission of radiation and energy transfer with the gas.
With the updated chemical abundances, 
we compute $\Gamma$ and $\Lambda$ for solving the energy equation.
The details of them are shown in \citet{Fukushima2021}.
We also include dust sublimation. 
We set the temperature of dust sublimation as $10^3~\rm K$, above which we ignore the contributions of dust grains on radiation, force attenuation, and thermal processes of gas components.

\begin{figure}
    \centering
	\includegraphics[width=\columnwidth]{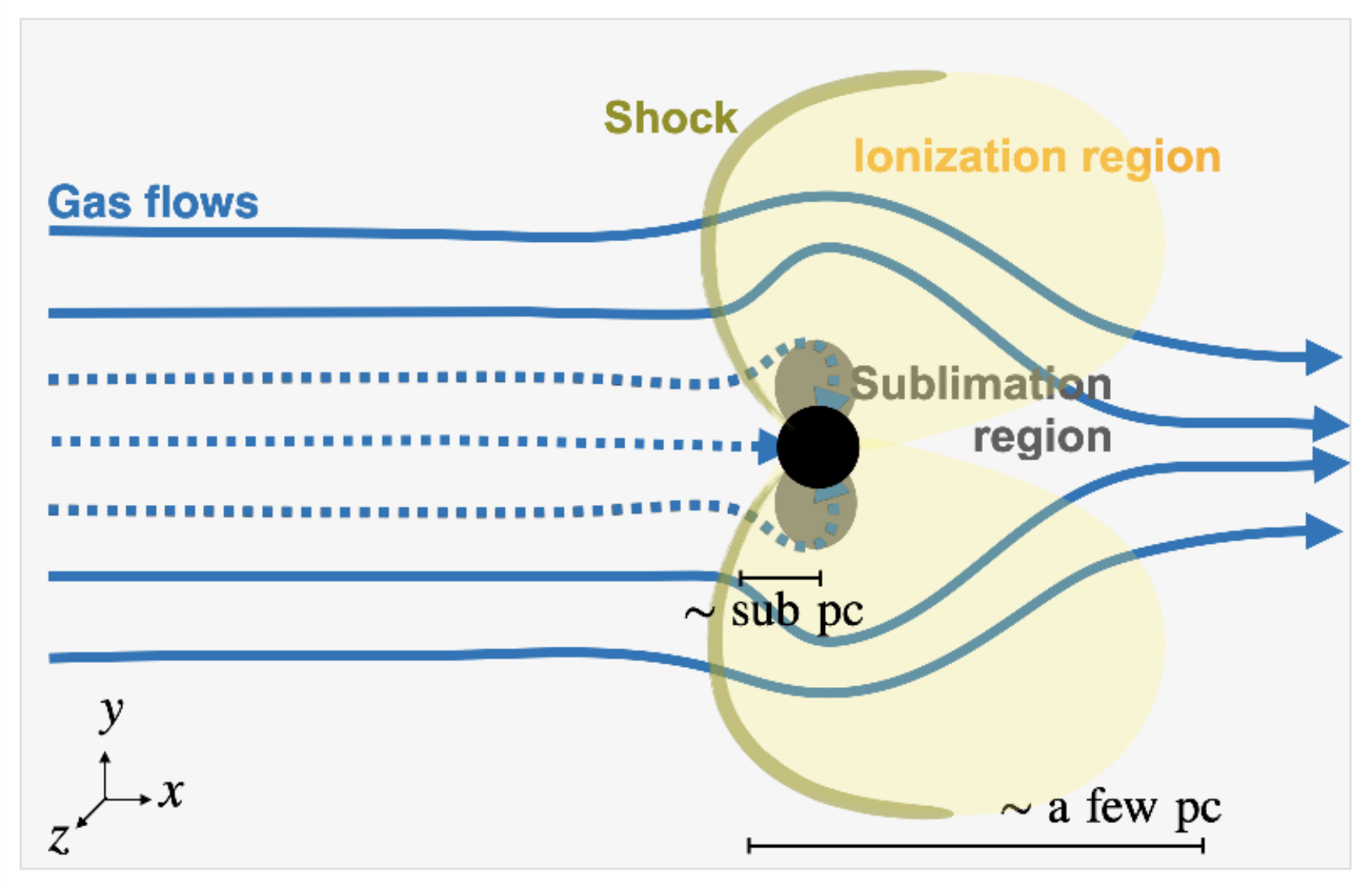}
    \caption{
    A schematic picture of a gas flow around a BH.
    We show a case with an edge-on radiation field.
    The BH floats in the interstellar medium (light gray).
    The black-filled circle represents the sink region.
    The solid and dotted blue lines show streamlines that pass through or accrete to BH.
    The yellow and gray regions present ionized and dust sublimation regions, respectively.
    The shocked region at the ionization front is painted dark yellow.
    The shapes of ionized and dust sublimation regions are different in the models with isotropic and pole-on radiation fields.
    }
    \label{fig:overview}
\end{figure}

\subsection{Radiative Source}
\label{S:Radiative Feedback}
 In our simulations, the BH luminosity $L$ is estimated with the mass accretion rate $\dot M$.
%We estimate the BH luminosity $L$ with the mass accretion rate $\dot M$.
We adapt the model of \citet{Watarai2000} as
\begin{equation}
    L = \left\{
    \begin{array}{ll}
    2L_{\rm E}[1+\ln{\Big( \frac{\dot{M}}{2\dot{M}_{\rm E}}\Big)}] & (\dot{M}/\dot{M}_{\rm E} > 2)\\
    L_{\rm E}\frac{\dot{M}}{\dot{M}_{\rm E}}& (\rm otherwise) 
    \end{array}
    \right.
    ,
    \label{eq:lum_watarai}
\end{equation}
where $L_{\rm E}~(\equiv4\pi c GM_{\rm BH}/\kappa_{\rm es})$ and $\dot{M}_{\rm E}~(\equiv L_{\rm E}/\eta c^2)$
are the Eddington luminosity and Eddington accretion rate. % defined with electron scattering.
Here, $M_{\rm BH}$, $\kappa_{\rm es}=0.4~\rm cm^2~g^{-1}$ and $\eta=0.1$ are the BH mass, opacity of electron scattering, and the radiative efficiency.
we assume that the radiation is emitted from the sink region with a single power-law spectrum $L_{\rm \nu} \propto \nu^{-\alpha}$ in the frequency range $11.2~{\rm eV} < h\nu < 1~{\rm keV}$, where $h$ and $\nu$ are planck constant and photon frequency, respectively.
Here, 
We adopt the power-law index of $\alpha=1.5$, corresponding to the spectral energy distribution of active galactic nuclei in the high accretion state \citep{Sazonov2004}.

We inject the extreme EUV and FUV photons at the sink region with the directional dependence as described below.
The number of photons injected per unit time in a single cell is given as
\begin{align}
    \dot{N}_{\nu}^j
    =
    \dot{N}_{\nu}^{\rm sink}\times 
    \frac{\mathcal{F}_j(\Theta, R)}{\sum_k \mathcal{F}_k(\Theta, R)}, 
    \label{eq_theta_dependece_of_irradiation}
\end{align}
where $\dot{N}_{\nu}^{\rm sink}$ is the photon emissivity of the BH, and $k$ and $j$ are the cell numbers. 
The anisotopy factor $\mathcal{F}_j(\Theta, R)$ represents the direction dependence of the radiation as 
\begin{equation}
    \mathcal{F}_j(\Theta, R)
    = 
    \left\{
    \begin{array}{ll}
    1/R^2 & (\rm isotropic ~radiation)\\
    2\cos{\Theta}/R^2 & (\rm anisotropic~ radiation) 
    \end{array}
    \right.
    .
    \label{eq:disk_rad}
\end{equation}
Here, the angle from the rotation axis of the disk to the center of each cell ($\Theta$) is defined as 
\begin{align}
    \cos{\Theta}
    &=
    \boldsymbol{i} \cdot \boldsymbol{s}
    =
    \frac{1}{R}
    \left(
    x\cos{\psi}+y\sin{\psi}\cos{\varphi}+z\sin{\psi}\sin{\varphi}
    \right),
    \\
%In Cartesian coordinates, 
%the unit vector $\boldsymbol{i}$ and the unit vector $\boldsymbol{s}$ are respectively
    \boldsymbol{i}
    &=
    (\cos{\psi},\sin{\psi}\cos{\varphi},\sin{\psi}\sin{\varphi}),
    \label{equation:rotation angle}
    \\
    \boldsymbol{s}
    &=
    \frac{1}{R}\left(
    x,y,z
    \right), \label{equation:vector_s}
\end{align}
where $R(=\sqrt{x^2+y^2+z^2})$ is the distance of the origin, 
$\varphi$ is the angle between $y$-axis and the disk rotation axis projected onto $yz$-plane,
and $\psi$ is the angle between $x$-axis and the disk 
rotation axis, respectively.
We set $\varphi=0$ in the present simulations.
We also solve the diffuse IR photons produced as the dust thermal emission.

\subsection{Cases examined}
\label{S:cases examined}

In our simulations, we fix the initial BH mass and the metallicity as $M_{\rm BH} = 10^4~M_{\odot}$ and $Z=0.1~Z_{\odot}$.
The simulation parameters are summarized in Table \ref{tab:model}.
Hereafter, we label each model according to the anisotropy factor ($\mathcal{F}_j$), the number density ($n_{\infty}$), and gas velocity ($v_{\infty}$) considered in each simulation.
For example, 'isoN4V20' represents the simulation model with the isotropic radiation field, $n_{\infty}=10^4\rm ~cm^{-3}$, and $v_{\infty}=20\rm ~km~s^{-1}$.

In the present study, $\psi=\pi/2$ (edge-on) is mainly adopted. 
The gas inflow direction ($-x$ direction) at the upstream boundary is perpendicular to the rotation axis of the disk in this model. 
Such a situation appears naturally. 
The angular momentum vector of the gas injected from the upstream boundary relative to the origin (BH) is perpendicular to the $x$-axis. 
Therefore, although the gas density at the upstream boundary is assumed to be uniform in our simulations, if the density is non-uniform, the total angular momentum vector of the gas will be perpendicular to the $x$-axis. 
Thus, the gas falling into the central region would form the accretion disk with the rotation axis perpendicular to the $x$-axis. 
Even if the rotation axis of the accretion disk is not perpendicular to the $x$-axis at the beginning, it should eventually become perpendicular to the $x$-axis since the gas in the initial disk is swallowed by the BHs, and the supplied gas becomes main component of the disk. 
The pole-on situation ($\psi=0$) is realized only at the moment if the black hole happens to enter the gas cloud from the direction of the rotation axis of the disk, but it is adopted in order to compare with the edge-on models.
In addition, for comparison with the model of anisotropic radiation, we perform the simulation with isotropic radiation.

We adopt the model with the gas number density $n_{\infty}=10^4\rm ~cm^{-3}$ and velocity $v_{\infty}=20\rm ~km~s^{-1}$ as the fiducial model.
We additionally investigate the cases with high-velocity $v_{\infty}=100~\rm km~s^{-1}$ and high-density $n_{\infty}=10^6~\rm cm^{-3}$ only in the models with edge-on anisotropic radiation.
Such high-density environments would appear, for example, in merger remnant galaxies \citep[e.g.][]{Fiacconi2013,Lima2017}. 
Also, \cite{Katz2023} has been reported the presence of relatively dense gas, $\gtrsim 10^4 \rm cm^{-3}$. 
The high-velocity situation, $v_\infty \sim 100~\rm km~s^{-1}$, might be realized 
when galaxy mergers occur \citep[e.g.][]{Mayer2007}.

%===================================================================================
%                                     RESULT
%===================================================================================

\section{Results}
\label{S:Results}

\subsection{Overview}
\label{S:Overview}

\begin{figure}
    \centering
	\includegraphics[width=\columnwidth]{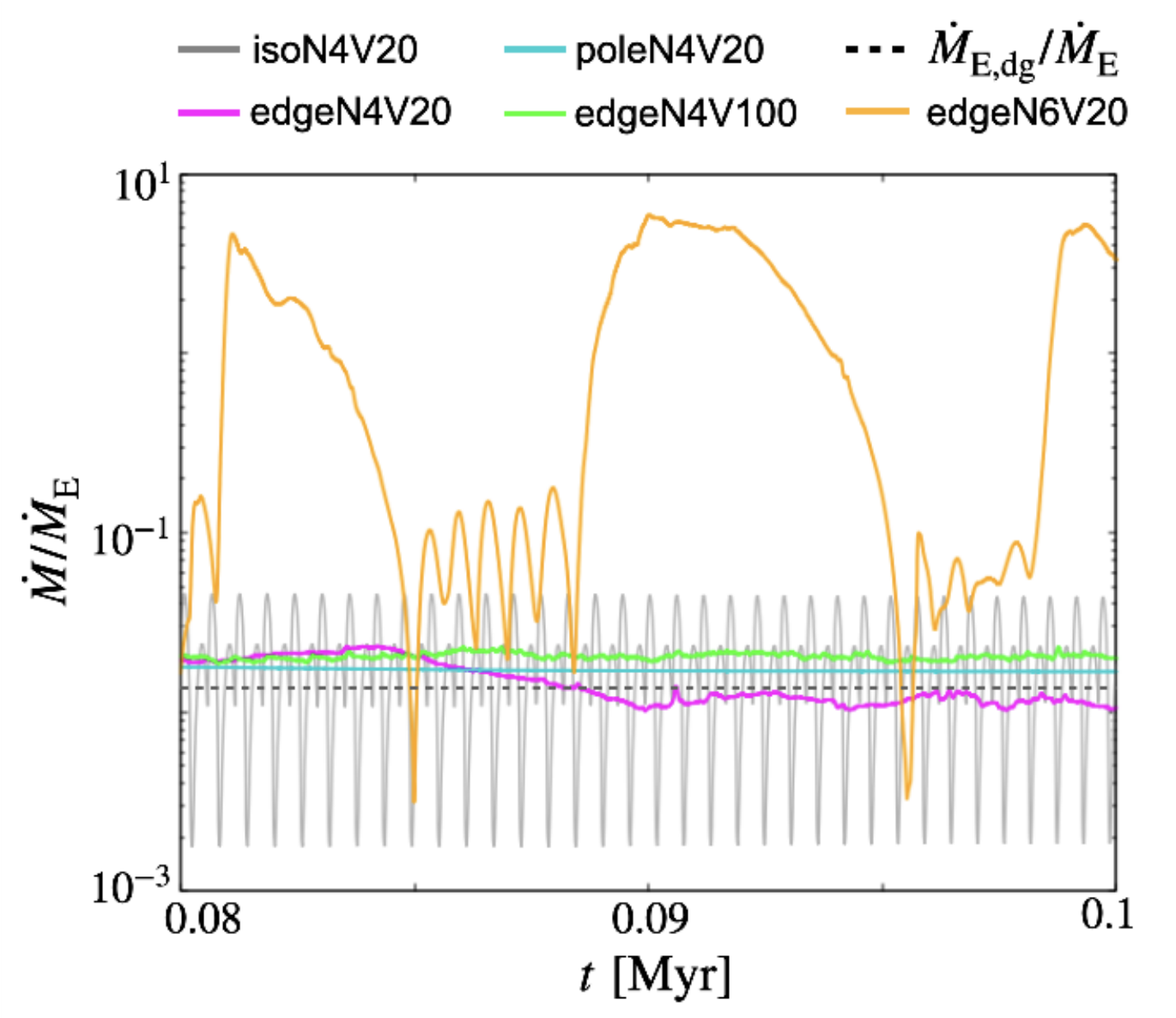}
    \caption{
    Time evolution of mass accretion rates.
    The vertical axis is normalized by the Eddington accretion rate $\dot{M}_{\rm E}$ estimated with the opacity of electron scattering.
    The gray and cyan lines show the models with isotropic radiation ('isoN4V20') and the pole-on radiation ('poleN4V20').
    The magenta, green, and orange lines represent the edge-on fiducial ('edgeN4V20'), high-velocity ('edgeN4V100'), and high-density models ('edgeN6V20').
    The black dotted lines represent the Eddington accretion rate for dusty gas $\dot{M}_{\rm E,dg}$.
    }
    \label{fig:mdot}
\end{figure}

Before examining the numerical results in more detail, we overview the flow structure and the accretion rates to BHs.
Figure \ref{fig:overview} schematically shows a flow structure around a BH.
Ionizing photons emitted from the accreting BH create an ionized region over several pc.
The shock structure is formed upstream ($-x$ direction) if the relative velocity of the ambient gas to the ionization front is between the critical values of the D-type and R-type ionization fronts ($v_{\rm D}<v_{\infty}<v_{\rm R}$), as also shown in \citet{Park2013}.
Inside the shocked region, the gas slows to $v_{\rm D}$ at the ionization front.
If the velocity $v_{\infty}$ is higher than the critical value for the R-type ionization front $v_{\rm R}$, the shock does not appear.
Dust grains are heated and sublimate close to the BH.
In this region, radiation force cannot push out the gas, and the gas tends to accrete to the BH.
In the outer region, 
where the dust grains are not sublimated, the gas is accelerated by radiation force and thermal pressure. 
Then it tends to pass through the ionized region without falling into the BH.

\begin{figure}
    \centering
	\includegraphics[width=\columnwidth]{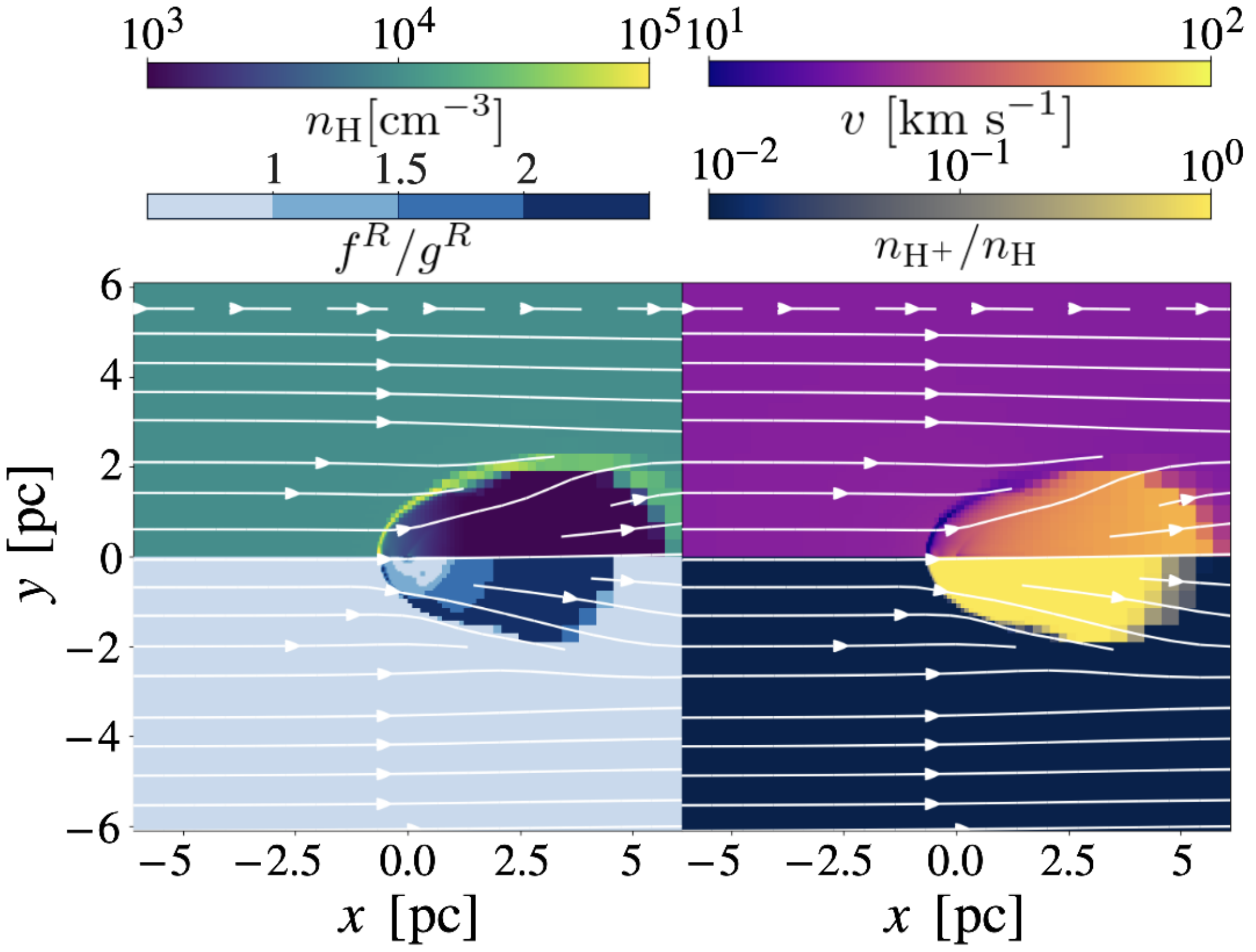}
    \caption{
    Gas flow structure on a $1~\rm pc$ scale for the isotropic radiation model ('isoN4V20').
    Each panel shows distributions of the gas number density (upper left), velocity (upper right), and ionization degree (lower right). 
    In the lower-left panel, we show the radiation force normalized by gravity in each cell. 
    The white arrows represent streamlines to the $+x$ direction.
    }
    \label{fig:iso_1pc}
\end{figure}

\begin{figure*}
    \centering
	\includegraphics[width=14cm]{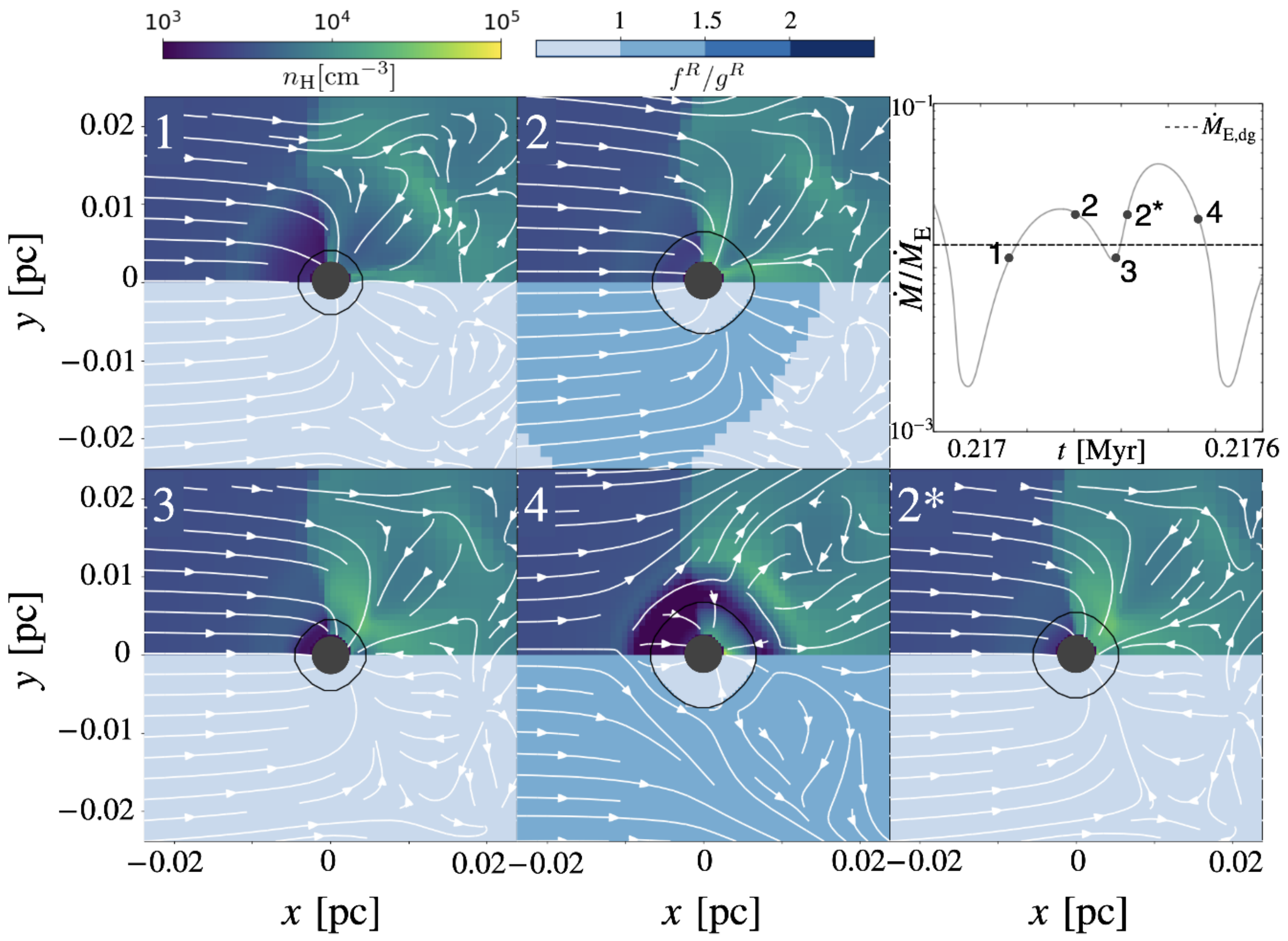}
    \caption{
    Time evolution of the gas flow in one oscillation period for the isotropic radiation model ('isoN4V20').
    Each snapshot is the same as the bottom left panel in Fig. \ref{fig:iso_1pc}.
    The black solid lines and filled gray circles represent the dust sublimation and sink regions.
    The upper right panel shows the time evolution of the accretion rate in one period of oscillation.
    The black filled circles mark the epochs for each snapshot 1-4. 
    }
    \label{fig:iso_peak}
\end{figure*}

Figure \ref{fig:mdot} shows the time evolution of the mass accretion rate. %s in each model. 
The black dashed line means the Eddington accretion rate for dusty gas $\dot{M}_{\rm E,dg}~(\equiv \kappa_{\rm es}\dot{M}_{\rm E}/(\kappa_{\rm es}+\kappa_{\rm d}))$,
where the dust opacity for UV light $\kappa_{\rm d}$ is given as $\kappa_{\rm d}=2.8\times 10^2 (Z/Z_{\odot})~\rm cm^2~g^{-1}$ \citep[e.g.][]{Yajima2017}.
The time-averaged accretion rates $\overline{{\dot{M}}}$ and luminosity $\overline{L}$ obtained in our simulations are summarized in Table \ref{tab:model}.
Here, we assume that the luminosity $L$ is the function of the accretion rate $\dot{M}$ (see Eq.\ref{eq:lum_watarai}).
In the isotropic radiation model 'isoN4V20', 
the accretion rate oscillates periodically between $4.5\times 10^{-2}\dot{M}_{\rm E}$ and $1.8\times 10^{-3}\dot{M}_{\rm E}$ with a period of $5.5\times 10^{-4}~\rm Myr$. 
The time-averaged accretion rate is about
$2\times 10^{-2}\dot{M}_{\rm E}$
($4.3 \times 10^{-6}~{M_{\odot}{\rm yr^{-1}}}$).
The accretion bursts (rapid increase in accretion rate) occur in the model 'edgeN6V20' with a period of $\sim 0.01~{\rm Myr}$.
The time-averaged accretion rate is about $2\dot{M}_{\rm E}$ ($4.2\times 10^{-4} M_{\odot} {\rm yr^{-1}}$).
Although it does not appear in the figure, model 'edgeN4V20' also exhibits accretion bursts at the interval with $\sim 0.15~{\rm Myr}$
(we will discuss later).
The time-averaged value in this model, $2 \times 10^{-2} \dot{M}_{\rm E}$ ($4 \times 10^{-6} ~M_{\odot} {\rm yr^{-1}}$, see Table \ref{tab:model}), is slightly larger than the accretion rate shown in Fig. \ref{fig:mdot} because the bursts increase the time-averaged rate.
The quasi-steady state is achieved in the models 'edgeN4V100' and 'poleN4V20', and the time-averaged rate is almost the same as in model 'edgeN4V20'.

\subsection{Isotropic case} \label{S:Isotropic Radiation model} 
Figure \ref{fig:iso_1pc} shows a flow structure of model 'isoN4V20' around the BH.
We find that an ionized region ($n_{\rm H+}/n_{\rm H} \sim 1$) extends up to $x \sim 5 ~{\rm pc}$, and a dense shock shell is formed at the upstream ionization front ($x \sim -0.5$pc).
The gas velocity gradually increases to $\sim 50~{\rm km~s^{-1}}$ in the ionized region after passing through the dense shock shell.
Such acceleration is caused by the thermal pressure increased by ionization heating.
Increasing the gas velocity and increasing the gas temperature reduce the accretion rate
since the Bondi-Hoyle-Lyttleton accretion rate decreases when the velocity and temperature are high.
The outward radiation force also woks to reduce the accretion rate. 
Thus, the time-averaged accretion rate, $2\times 10^{-2}\dot{M}_{\rm E}$, is three orders of magnitude smaller than the classical Bondi-Hoyle-Lyttleton accretion rate $\dot{M}_{\rm BHL}$. 
We will show the comparison with previous work \cite{Toyouchi2020} that does not include the dust sublimation in Section \ref{S:Previous studies}.

The reason why the periodic oscillation of the accretion rate occurs can be understood from Fig. \ref{fig:iso_peak}. This figure shows the time evolution of flow structure and the radiation force normalized by gravity on the $10^{-2}~\rm pc$ scale in one period of oscillation.
The time evolution of the accretion rate is also plotted.
When the gas accretion rate is less than $\dot{M}_{\rm E,dg}$, the gravity overcomes the radiation force in the entire region (Fig.\ref{fig:iso_peak}-1). 
The gas moves towards the BH, and the flow structure becomes similar to Bondi-Hoyle-Lyttleton accretion.
When the accretion rate exceeds $\dot{M}_{\rm E,dg}$ (Fig. \ref{fig:iso_peak}-2), the radiation force reduces the accretion rate.
The reason why gas accretion does not stop completely is because the radiation force is not effective in the region where dust sublimes near the BH (inside the black solid lines).
When the accretion rate decreases to $\dot{M}_{\rm E,dg}$, gravity overcomes radiation force. Thus, the accretion rate increases again (Fig.\ref{fig:iso_peak}-3) and then begins to decrease (Fig.\ref{fig:iso_peak}-4) due to the strong radiation force.

\subsection{Edge-on case}

\subsubsection{Fiducial model}\label{S:Anisotropic Radiation model}

\begin{figure*}
    \centering
	\includegraphics[width=17cm]{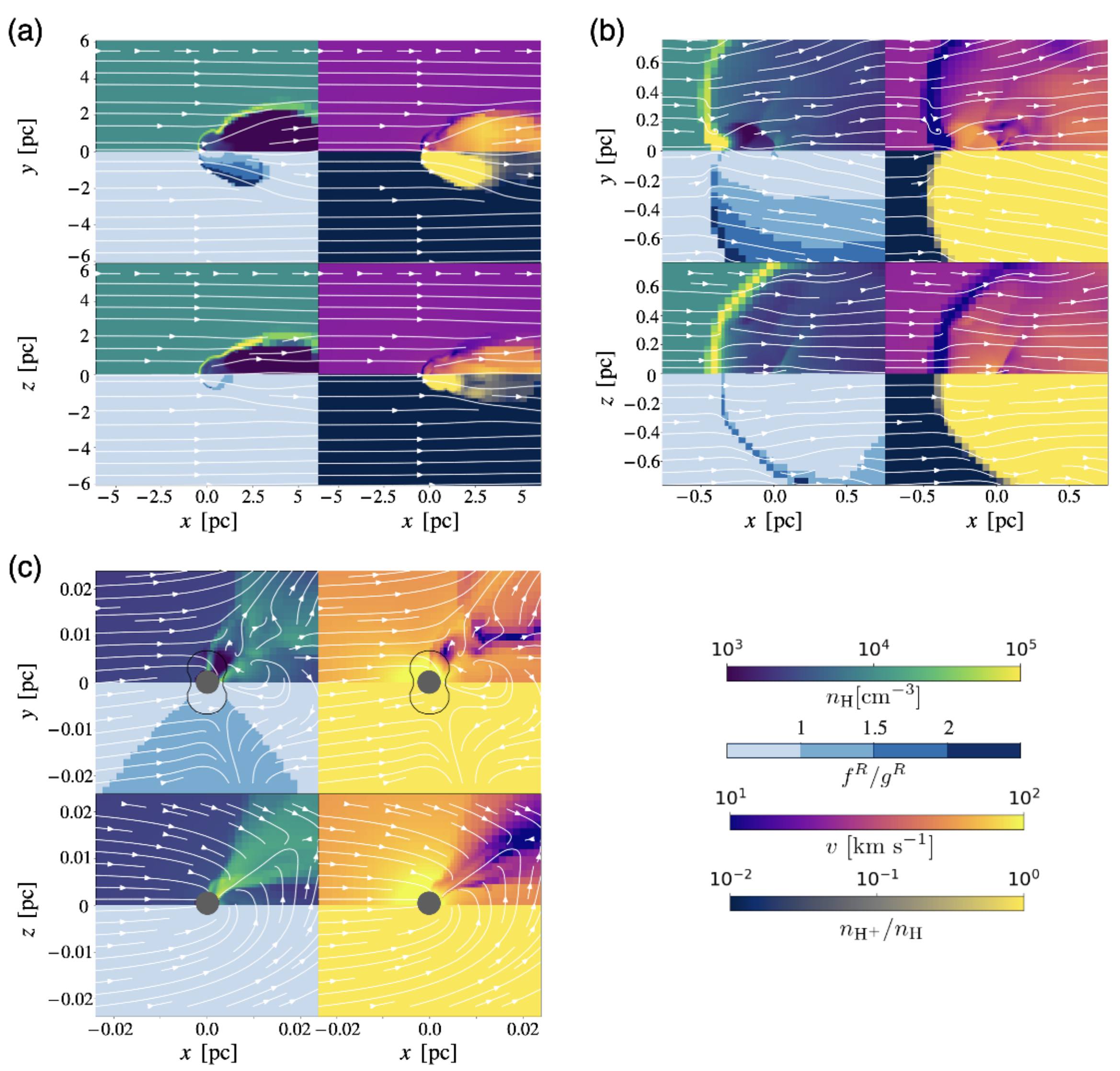}
    \caption{
    Same as Fig. \ref{fig:iso_1pc} but for the edge-on fiducial model ('edgeN4V20'). 
    Each panel shows the flow structure on (a)$1~\rm pc$, (b)$10^{-1}$ and (c)$10^{-2}~\rm pc$ scales at $t=0.2~\rm Myr$ ($0.04~\rm Myr$ before the next accretion burst).
    Each column represents the flow structure in the meridian (xy-) and equatorial (xz-) planes from top to bottom.
    }\label{fig:edge_flow}
\end{figure*}

In the case of 'edgeN4V20' (fiducial model),
the accretion rate is almost constant for most of the time (quiescent phase), but accretion bursts occur periodically (burst phase).
Figure \ref{fig:edge_flow} presents the flow structure in the meridian ($z=0$) and equatorial ($y=0$) planes in the quiescent phase.
Although the radiation field is anisotropic, the profiles of density and velocity on the $\sim 1~{\rm pc}$ scale are almost similar to that of the isotropic radiation model (see Fig. \ref{fig:edge_flow}-a).
Also in this model, an ionized region extends up to several pc. 
Around the surface of the dense shock shell, the gas pressure gradient force by the high density and high temperature gas in the shell moves the gas away from the BH.
Indeed, we find that some gas moves away from the $x$-axis along the shell (see streamlines).
The effect of anisotropic radiation can be seen in Fig. \ref{fig:edge_flow}-b.
It is found that the shape of the ionization region is like two spheres glued together at the equatorial plane. 
The dense shock shell appears near the ionization front.
The velocity and temperature of the gas that passes through the shell and enters the ionized region rapidly increases by gas pressure gradient force and the ionization heating.
These increases in velocity and temperature work to reduce the rate of accretion by the Bondi-Hoyle-Lyttleton mechanism.
We find in Fig. \ref{fig:edge_flow}-c that
the gas accretes mainly from the equatorial plane since the radiation force is ineffective because BH irradiation is considerably weakened by absorption. 
In contrast, the strong radiation force works to prevent gas accretion around the rotation axis of the disk.
As a result, the time-averaged accretion rate in the quiescent phase is much smaller than Bondi-Hoyle-Lyttleton rate ($\sim 0.3\%$ of $\dot{M}_{\rm BHL}$), and nearly comparable to $\dot{M}_{\rm dg}$.

The accretion bursts only increase the time-averaged accretion rate by about a factor of 2, and also hardly affect acceleration at all (we will discuss later).
The bursts occur when part of the dense shock shell (near the $x$-axis) flows into the sink. 
The dense shock shell is almost never moved due to the balance between the ram pressure gradient force and the gas pressure gradient force in the quiescent phase. 
However, part of the dense shock shell around the $x$-axis gradually moves inward due to the gravity of the BH and then flows into the sink leading to the burst phase. 
The accretion rate of the burst phase is $\sim \dot{M}_{\rm E}$ and about half of the
total accreting gas accretes during the burst phase. 
Due to the bursts, the time-averaged accretion rate (Table \ref{tab:model}) is approximately twice the accretion rate during the quiescent phase (see Fig. \ref{fig:mdot}).

The burst interval, $\sim 0.15\rm Myr$, can be understood approximately as follows. 
In the situation where gas pressure and ram pressure are approximately balanced at the dense shock shell, the gravity begins to pull the shell when the column number density of the shell
($N_{\rm shell}$) exceeds $n_{\infty} r_{\rm shell}^2 v_{\infty}^2/GM_{\rm BH}$ because the gravity, $GM_{\rm BH}N_{\rm shell}/r_{\rm shell}^2$, becomes larger than the ram pressure (gas pressure), $n_{\infty} v_{\infty}^2$.
Since the distance of the dense shock shell from the BH, $r_{\rm shell}$, is about $0.5\rm pc$, the critical column number density obtained from the above relation is $\sim 10^{23} \rm cm^{-2}$, which is roughly consistent with that the burst in the present simulation.
The timescale on which the column number density of the dense shock shell reaches $10^{23} \rm cm^{-2}$ is approximately $\left(n_{\infty} r_{\rm shell}^2 v_{\infty}^2/GM_{\rm BH}\right)/(n_{\infty} v_{\infty})\sim v_{\infty} r_{\rm shell}^2/GM_{\rm BH}$.
This timescale is evaluated as $\sim 1.1 \rm Myr$ and is longer than the free-fall time so the burst interval is determined by this timescale.

\subsubsection{Higher velocity model}
\label{S:Higher Velocity Case}

\begin{figure*}
    \centering
	\includegraphics[width=17.5cm]{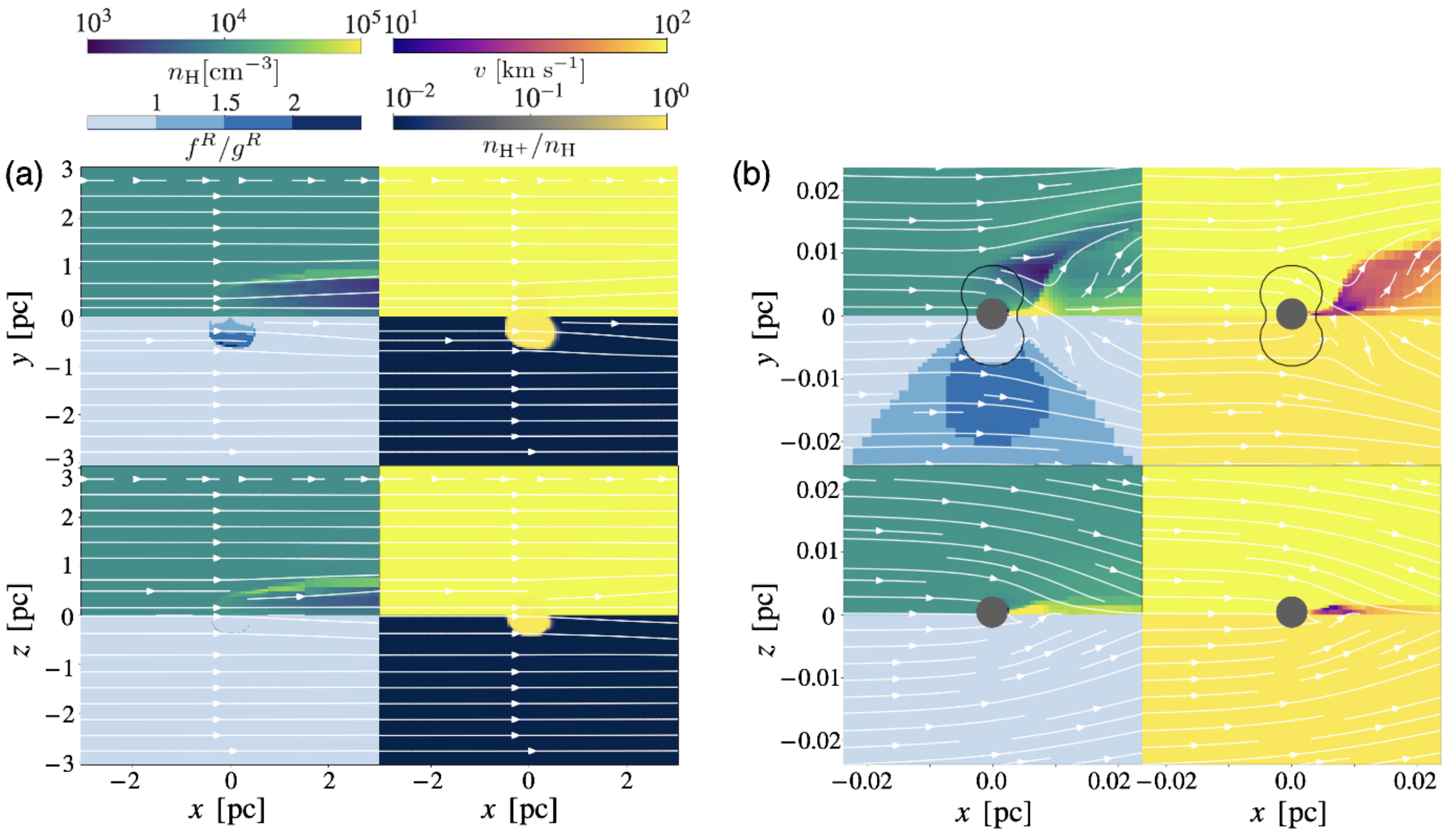}
    \caption{
    Same as Fig.\ref{fig:edge_flow} but for the edge-on high-velocity model ('edgeN4V100').
    Each panel shows the flow structure on (a) $1~{\rm pc}$ and (b) $10^{-2}~{\rm pc}$ scales.
    The black solid lines in panel (b) represent the dust sublimation regions. 
    }
    \label{fig:edge_v_flow}
\end{figure*}

Figure \ref{fig:edge_v_flow} shows the flow structure in the model 'edgeN4V100'.
The most noticeable difference from the fiducial model is the absence of shock waves.
This is because the relative velocity is largar than $v_{\rm R}$.
We also confirm that the density does not change so much at the ionization front. 
The density change at the ionization front can be estimated analytically with the mass and momentum conservation laws as ${\rho_{\rm HII}}/{\rho_{\infty}} \sim  {v_{\infty}^2} \left( 1- \sqrt{1-{4c_{\rm s,HII}^2}/{v_{\infty}^2}} \right)/{2c_{\rm s,HII}^2}$ under the condition $c_{\rm s,HII} \gg c_{\infty}$,
where $\rho_{\rm HII}$, $c_{\rm s,HII}$, and $ c_{\infty}$ are the gas density of ionized gas and sound speeds of ionized and neutral gas \citep{Spitzer1978}.
If $v_{\infty} \gg c_{\rm s,HII}$, then we obtain ${\rho_{\rm HII}} \sim {\rho_{\infty}}$.
We confirm in our simulations that the condition of $v_{\infty} \gg c_{\rm s,HII}$ is satisfied.
The somewhat low-density region appears over several pc downstream ($x>0$, $y\lesssim 1$ pc, dark-purple in Fig. \ref{fig:edge_v_flow}-a). 
This is because the gas is accelerated by radiation force and gas pressure gradient force in the ionized region.

The radiation force does not affect the the motion of gas at distances within the Bondi-Hoyle-Lyttleton radius from the $x$-axis
so that the flow structure is similar to the classical Bondi Hoyle-Littleton accretion.
Figure \ref{fig:edge_v_flow}-(b) shows the flow structure in the $10^{-2}~{\rm pc}$ scale.
Around the disk rotation axis, the radiation force is basically stronger than gravity, but only near the BH is it weaker than the gravity due to sublimation of the dust grains.
Since the Bondi-Hoyle-Lyttleton radius is around $9\times 10^{-3}$ pc, and since the dust sublimation radius around the rotation axis is $\sim 7\times 10^{-3} {\rm pc}$, the radiation force does not affect the motion of gas at distances within the Bondi-Hoyle-Lyttleton radius from the $x$-axis.
Thus, a quasi-steady flow similar to Bondi Hoyle-Littleton accretion appears.
The accretion rate is roughly comparable to $\dot{M}_{\rm BHL}$ (see Table \ref{tab:model}).

\subsubsection{Dense model} 
\label{S:Dense Environment Case}

\begin{figure}
    \centering
	\includegraphics[width=\columnwidth]{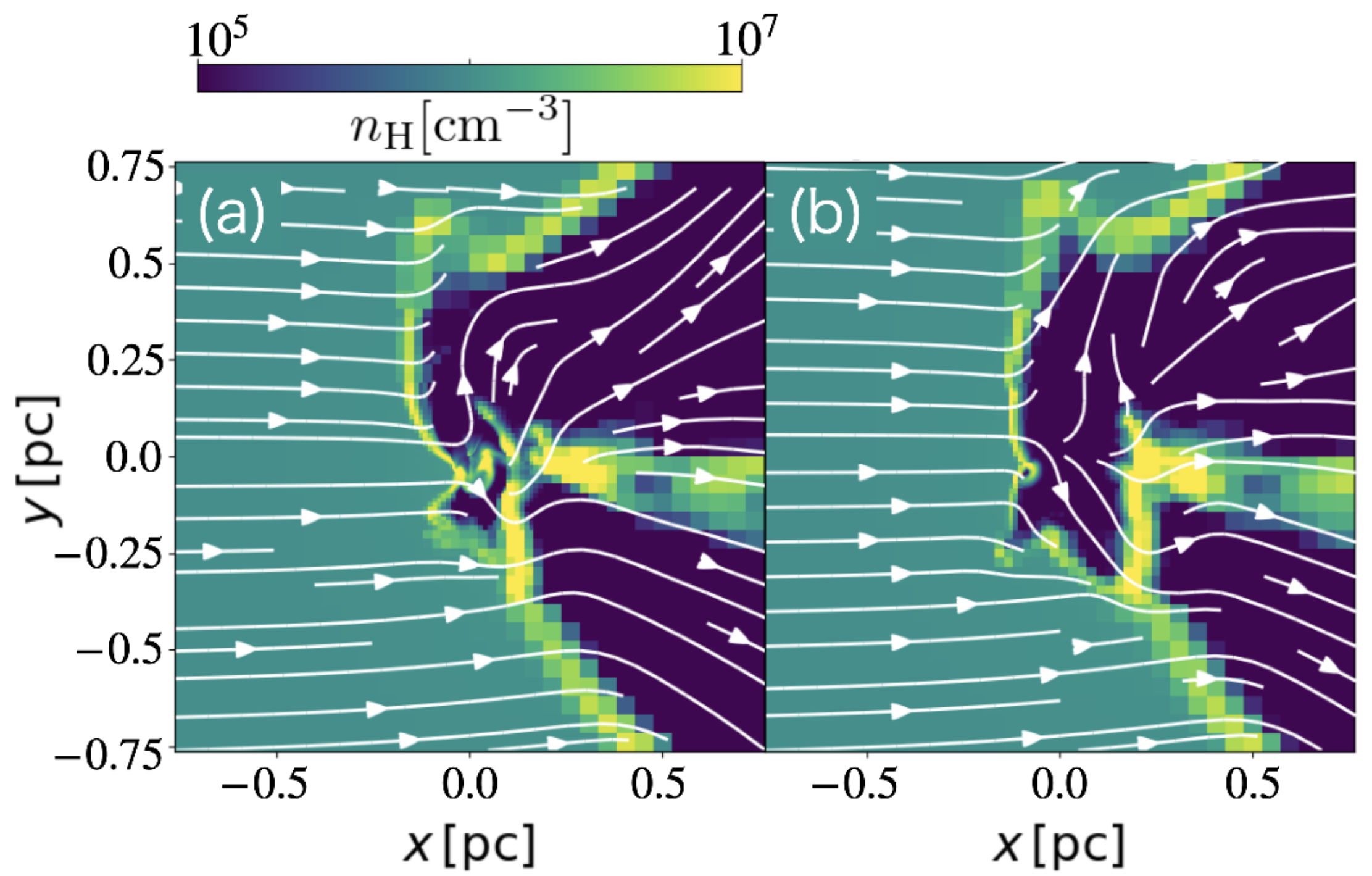}
    \caption{
    The flow structure on $10^{-1}~\rm pc$ scale for the edge-on high-density model ('edgeN6V20').
    Each panel shows the number density distribution in the equatorial (xy-) plane at (a) the burst and (b) quiescent phases.
    We also show the mass accretion rates and the elapsed times in each panel.
    }
    \label{fig:edge_hn_nH}
\end{figure}

Model 'edgeN6V20' differs from the fiducial model in that it produces more intense, shorter-interval accretion bursts.
Figure \ref{fig:edge_hn_nH} shows the time evolution of the gas density distribution in the meridian plane ($z=0$) for the model 'edgeN6V20'.
As shown in Fig. \ref{fig:mdot}, the accretion bursts occur periodically in this model.
The burst mechanism is similar to that of fiducial model.
In the quiescent phase, the dense shock shell exists near the ionization front, and it gradually becomes denser.
After a while, the gravity causes part of the shell to reach the sink, and the burst phase begins (see Fig.\ref{fig:edge_hn_nH}-a).
In the burst phase, the accretion rate exceeds the Eddington limit. 
However, the radiation force is ineffective around the equatorial plane due to the attenuation of radiation and the thermal pressure of inoized bubbles cannot push outward these dense gas.
After the significant part of the dense shock shell falls into the BH through the vicinity of the equatorial plane,
it transitions to the quiescent phase (see Fig.\ref{fig:edge_hn_nH}-b).
After that, the part of the burst shell falls into the BH, and the burst accretion occurs again. 
The above process is repeated, resulting in periodic bursts.
The interval of the accretion burst, $\sim 0.01\rm Myr$, is much shorter than that of the fiducial model, $\sim 0.15\rm Myr$. 
This is because the $r_{\rm shell}$ in this model, $\sim 0.1$ pc, is smaller than that of the fiducial model (it was shown in section \ref{S:Anisotropic Radiation model} that the burst interval depends on the $r_{\rm shell}$).

At the burst phase, the radius of the ionized region, $R_{\rm HII}$, is comparable to or slightly smaller than the Bondi-Hoyle-Lyttleton radius, $R_{\rm BHL}$, at around the equatorial plane where gas mainly accretes. 
On the other hand, we find $R_{\rm HII}>R_{\rm BHL}$ in regions other than the equatorial plane. 
This means that the condition for super-Eddington accretion "$R_{\rm HII}<R_{\rm BHL}$" \citep{Inayoshi2016} is satisfied locally. 
The effect of dust, which increases the radiation force and narrows the ionized region, on the condition for super-Eddington accretion should be investigated in detail in the future.

%========================pole on==============================%
\subsection{Pole-on case}
\label{S:Structures of flows; pole-on}

\begin{figure}
    \centering
	\includegraphics[width=\columnwidth]{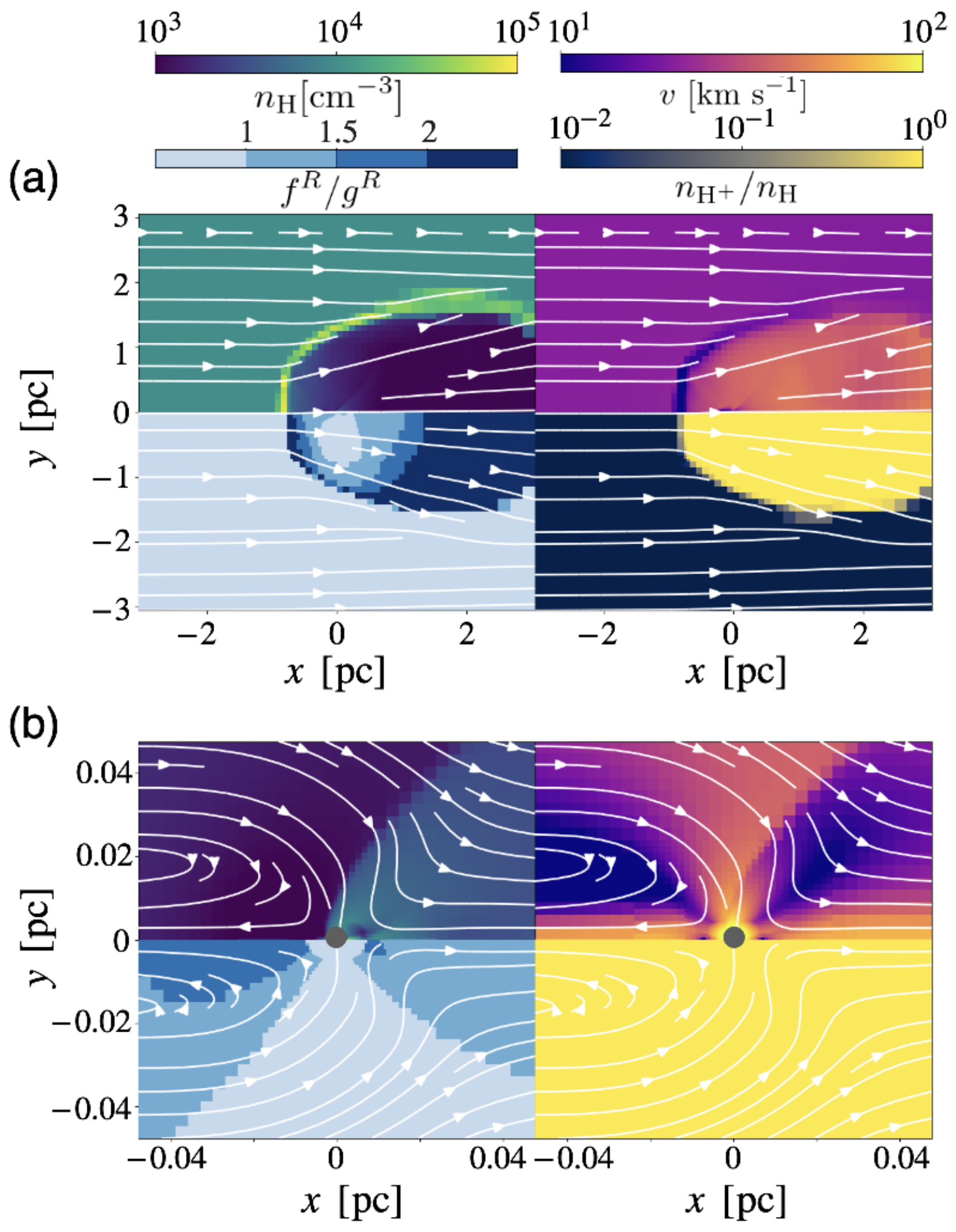}
    \caption{
    Same as Fig. \ref{fig:edge_flow} but for the pole-on model ('poleN4V20').
    Each panel shows the flow structure on (a) $1~\rm pc$ and (b) $10^{-2}~\rm pc$ scales. 
    We only show the snapshot in the xy-plane due to the axisymmetry of flow.
    }
    \label{fig:pole_flow}
\end{figure}

Figures \ref{fig:pole_flow}-(a) and -(b) show the flow structure on a 1 pc and $10^{-2}$ pc scales of the pole-on radiation model ('poleN4V20').
The flow structure is quasi-steady and axisymmetric with respect to the $x$-axis.
The structure on pc scale is almost the same as the fiducial model.
The ionized region extends to several pcs, and the dense shock shell is formed at the ionization front
(see Fig.\ref{fig:pole_flow}-(a)).
As shown in Fig. \ref{fig:pole_flow}-(b),
this model also shows that gas accretion occurs mainly from near the equatorial plane ($x=0$ plane),
where the gravity is stronger than the radiation force.
In addition, the gas that is attracted by the gravity but does not fall into the BH reaches near the $x$-axis at the downstream region and then flows out in the $+x$ direction.
Unlike the fiducial model, the vortex flows appear in the upstream region.
This is due to the radiation force pushing the gas in the $-x$ direction.

\section{Discussion}
\label{S:Discussions}
%{\color{red}<discussionは単数のはずですがチェック>}

\subsection{Evolution of IMBHs}
\label{S:Dynamical Friction}

\begin{figure}
    \centering
	\includegraphics[width=\columnwidth]{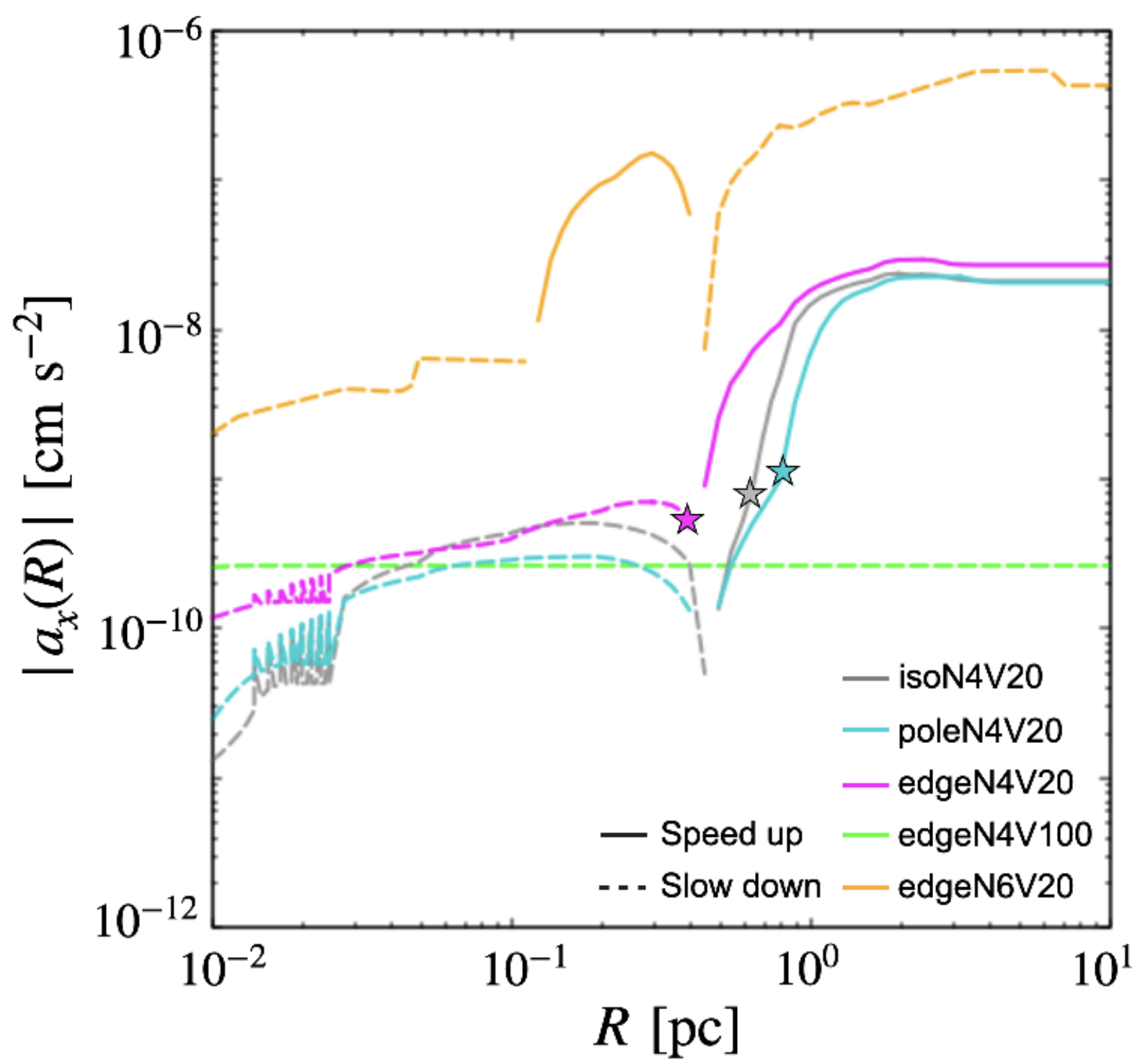}
    \caption{
    Acceleration in the $x$-direction of BHs at $t\sim 8.5\times 10^{-2}~\rm Myr$ as a function of $R$. 
    The solid (dotted) lines mean that the acceleration is negative (positive) and the speed of the BHs increases (decreases).
    The line colors are the same as in Fig. \ref{fig:mdot}.
    The star markers correspond to the positions of the ionization fronts. 
    }
    \label{fig:time_df}
\end{figure}

Gravity by non-uniform gas distribution around the BHs 
and momentum transport to the BHs by gas accretion induce acceleration of BHs.
Figure \ref{fig:time_df} shows the acceleration in the $x$-direction $a_x(R)$ at the elapsed time $t = 8.5\times 10^{-2}~\rm Myr$, which is calculated as 
\begin{equation}
    a_x(R) =\int_0^{4\pi} \int_{R_{\rm sink}}^{R} \frac{G\rho x}{R'^3} dR' d\Omega + \int_{S_{\rm sink}} \rho v_x \boldsymbol{v} \cdot d\boldsymbol{S}_{\rm sink},
\end{equation}
%{\color{red}<式が太字>}
where $d\Omega$ and $d\boldsymbol{S}_{\rm sink}$ are the solid angle and area vector.
%{\color{red}<単なる面積ではなくベクトルです. area vectorと書いたりするはずです>}
Here, a negative (positive) value of $a_x(R)$ means that the BH accelerates in the upstream (downstream) direction and the velocity of BH relative to the interstellar gas increases (decreases).

It is found that the acceleration $a_x (R_{\rm out})$ is about $-10^{-8}\rm cm~s^{-2}$ in the models with $v_{\infty} = 20~{\rm km~ s^{-1}}$ and $n_{\infty}=10^4~\rm cm^{-3}$
('isoN4V20', 'poleN4V20', and 'edgeN4V20').
This is mainly due to gravity from the dense shock shell on the ionization front. 
This can be understood from the fact that the acceleration suddenly increases
at around the star marks (the positions of the ionization front)
and becomes nearly constant outside of them.
In the model with the high-velocity $v_{\infty} = 100~{\rm km/s}$ ('edgeN4V100'), 
we find $a_x (R_{\rm out}) \sim 10^{-10}\rm cm~s^{-2}$, which is consistent with the result of \cite{Ostriker1999}, in which the acceleration of the central object of Bondi-Hoyle-Lyttleton flow has been evaluated.
Positive acceleration is due to the accretion of gas with positive momentum ($\rho v_x>0$) onto the BH.
Although the gas density outside the sink in the upstream region is higher than that in the downstream (see section \ref{S:Higher Velocity Case}),
the negative acceleration due to this density difference is smaller than the positive acceleration via the gas accretion.
In the high-density case ('edgeN6V20'), the acceleration in the burst phase is about $a_x (R_{\rm out})\sim 10^{-7}\rm cm~s^{-2}$, and we find the gravity of the dense downstream gas induces the positive acceleration.
Although the acceleration becomes negative due to the gravity of the dense shock shell in the quiescent phase, time-averaged acceleration is positive, $\sim 10^{-7}\rm cm~s^{-2}$.

In the following, we discuss the evolution of the mass and velocity of the IMBHs ($M_{\rm BH} \sim 10^4M_\odot$) floating in early galaxies ($z\gtrsim 6$)
using our results of model 'edgeN4V20', 'edgeN4V100', and 'edgeN6V20'.
Here, we use $a (R_{\rm out})$ at $t\sim 0.1\rm Myr$ presented in Fig. \ref{fig:time_df} instead of the time-averaged value since the acceleration at larger distance ($R\gtrsim$several pc) does not change with time as much.

In the case that the gas density is $\sim 10^4\rm ~cm^{-3}$, the IMBHs continue to float without significant mass growth.
For the model 'edgeN4V100', 
the acceleration is $\sim 10^{-10}\rm~cm~s^{-2}$ so that the timescale of the slowdown is $\sim 3~\rm Gyr$. This is much longer than the age of the universe at $z=6$ ($\sim 900~\rm Myr$).
This implies that the speed of IMBHs does not change so much.
On the other hand, 
the acceleration timescale of model 'edgeN4V20', $\sim 6~\rm Myr$, is much shorter than $900~\rm Myr$.
Thus, if the initial velocity is relatively small, $\sim 10 ~ \rm km~s^{-1}$, the IMBHs will continue to speed up until the velocity reaches ${\rm several} \times 10 ~ \rm km~s^{-1}$ where the acceleration timescale becomes comparable to the age of the universe.
The above discussion does not take into account the change in the mass of the IMBHs, but this is appropriate.
In both model 'edgeN4V100' and 'edgeN4V20', 
the timescale of mass growth derived from the accretion rate ($\sim 4\times 10^{-6}~M_{\odot}~\rm yr^{-1}$) is $\sim 2~\rm Gyr$, which is much longer than $900~\rm Myr$.
Thus, we conclude that the IMBHs continue to float at the velocity of $\gtrsim {\rm several} \times 10 ~ \rm km~s^{-1}$ without significant mass growth.

For higher gas number densities ($\sim 10^6~\rm cm^{-3}$), IMBHs would slow down and increase in mass.
Although we have not performed the simulations of high-density, high-velocity model ($n_\infty = 10^6~\rm cm^{-3}$ and $v_\infty=100 \rm km~s^{-1}$ ), 
the classical Bondi-Hoyle-Lyttleton type flow is expected to emerge because the flow structure in model 'edgeN4V100' is similar to the Bondi-Hoyle-Lyttleton accretion flow due to the small impact of radiation, and because the radiation impact is thought to be more ineffective as the density is increased.
Hence the accretion rate can be inferred from the classical Bondi-Hoyle-Lyttleton accretion theory as $\sim 6\times 10^{-4}M_{\odot}\rm yr^{-1}$,
and the acceleration is evaluated to be $-4\times 10^{-8}~\rm cm~s^{-2}$ based on the prediction of \cite{Ostriker1999}. 
The timescales of mass growth and slowdown of IMBHs are expected to be $\sim 20~\rm Myr$ and $\sim 7 ~\rm Myr$, respectively.  
This means that the relative velocity decrease faster than mass growth if the initial velocity is $\sim 100~\rm km~s^{-1}$.
This behavior is thought to continue even after the relative velocity drops to ${\rm a\,\,few} \times 10~\rm km~s^{-1}$.
This is because that the slowdown is expected to be more pronounced due to the decrese of the velocity, while the mass accretion rate remains the same.
Indeed, in the model 'edgeN6V20', the deceleration rate is $\sim 10^{-7}\rm~cm~s^{-2}$ and the timescale of the slowdown, $\sim  0.4 ~\rm Myr$, is much shorter than the high-velocity case. 
The accretion rate (timescale of the mass growth), for $v_\infty =20 \rm km~s^{-1}$ and $100 \rm km~s^{-1}$ is almost the same.
Subsequently, a shift from Bondi-Hoyle-Lyttleton accretion to Bondi accretion might occur.
Bondi accretion in an anisotropic radiation from accretion disks around BHs has been investigated by \cite{Sugimura2017}, and the accretion rate has been reported to be 1\% of the Bondi rate.
Adopting $n_\infty$ and $T_\infty$ as number density and temperature of the gas, 
the timescale of the mass growth is $16~\rm Myr$,
which is much shorter than the age of the universe.
To sum up, the IMBHs floating at speeds of $\sim 10-100\rm km~s^{-1}$ rapidly slow down and grow. Eventually supermassive BHs may be formed.

In the present discussion, we only consider the interaction between a single BH and the surrounding gas. 
To more accurately investigate the evolution of IMBHs, it should be necessary to consider their interactions with stars and other BHs.

%{\color{red} <速度20をfiducial modelとし、速度100もアリ得るという立場の論文で、速度100がもっともらしいと書くと自己矛盾する.>The BHs floating in the interstellar space are expected to encounter a gas cloud with a relative velocity of approximately $100~\rm km~s^{-1}$ \citep[e.g.][]{Mayer2007}.}

\subsection{Comparison with previous works}
\label{S:Previous studies}
Our present work demonstrates that the dust absorption significantly changes the flow structure around moving BHs.
In the low-velocity cases (model 'edgeN4V20' and 'edgeN6V20'),
the accretion rate is several times smaller than the analytical solution of \citet{Park2013},
in which dust-free situation is supposed.
One reason for the such a small accretion rate may be due to the radiation force acting on the dust \citep[see][]{Toyouchi2020}.
However, the presence of dust does not significantly affect the gas flow in the high-velocity case.
For the model 'edgeN4V100', the accretion rate is almost consistent with the analytical solution of \citet{Park2013} and classical Bondi-Hoyle-Lyttleton rate. 
This is because the radiation force does not change the flow structure, and a classical Bondi-Hoyle-Littleton type flow emerges.

It is also important to take into account  the dust sublimation.
The time-averaged accretion rate is not much different between our simulation 'isoN4V20' and the simulations by \cite{Toyouchi2020}, but oscillation of the accretion rate occurs only in our simulation 'isoN4V20'.
Since the situational setting is almost the same, the cause of this difference is thought to be the treatment of dust sublimation.
In addition, 
the result that the moving IMBH is accelerated by the gravity when the dense shock shell is formed is consistent with previous studies  \citep{Park2017, Toyouchi2020}, but the acceleration obtained with model 'isoN4V20' is slightly smaller than that of \citet{Toyouchi2020}.
This is because, in our simulations, the dense shock shell is formed somewhat further away from the IMBH.
The reason for this is expected to be the small sink size and the implementation of dust sublimation, but the details are to be worked out in the future.

Here, we compare the present study with our previous work that investigated the Bondi-Hoyle-Littleton accretion of dusty gas in an anisotropic radiation field without solving the hydrodynamics equations.
The accretion rate is 0.6\% of $\dot{M}_{\rm BHL}$ in the model 'edgeN4V20', while it is $20-30$\% for the model of the previous work that have the same luminosity as that of the quiescent phase of model 'edgeN4V20'.
In the model 'edgeN4V20', the gas is pushed out due to thermal pressure on the ionization front and a large amount of gas passes through without being swallowed by the BH.
Thus, the accretion rate is drastically reduced.
This means that hydrodynamics calculations are necessary, at least when shock is formed. 
Otherwise, the accretion rate would be overestimated.

\subsection{Future work}
\label{S:Effects neglected}

In this study, we assume the uniform density distribution of the ambient gas.
However, the interstellar medium is inhomogeneous, which could alter the acceleration rate and acceleration of BHs.
\cite{Ruffert1997, Ruffert1999} have reported that non-uniformity leads to flows orbiting around the sink, which reduces the accretion rate.
Recently, \cite{Lescaudron2022} suggested that turbulent medium decelerates BHs using the three-dimensional magneto-hydrodynamics simulations.
In their simulations, the deceleration rate increases by several tens of times that of the analytical solution with a uniform density \citep{Ostriker1999}.
However, the moving BHs could be accelerated if the radiation is taken into account.
Although \cite{Sugimura2018} has calculated the 
accretion of gas with angular momentum onto a static BH,
the simulations of moving BHs in the inhomogeneous gas, considering the radiation feedback, have not yet been performed.
These are important future work.

There would also be room to modify the handling of accretion disks around the BHs.
The rotation axis of the disk is likely to be perpendicular to the direction of gas motion  (edge-on) as described in Section \ref{S:cases examined}.
However, the gas accreting to 
the central region could have random angular momentum if the ambient gas is inhomogeneous.
In such a case, the edge-on disk is not maintained.
In addition, immediately after the BHs encounter the gas clouds, the situation is likely to be different from edge-on case.
In these cases, the shape and location of the dense shock shell and ionization front could be different from the results of the present work.
To be realistic, the improvements mentioned above are needed.

In our simulations, the gas flowing into the sink is assumed to accrete to the BH and change the luminosity immediately.
However, more precisely, the luminous intensity should increase with a delay of about the viscous time.
This time lag would not be negligible
in the case that the angular momentum of the accreting gas is large.
Clarifying these points is an important future work.

Spectral energy distribution (SED) of BHs depend on mass accretion rates.
In this study, we assume a power-law SED (see also Section \ref{S:Simulation Methods}).
If the mass accretion rate is near or above the Eddington rate, the SED of the accretion disk would be multi-temperature blackbody radiation \citep{Kato2008}.
Additionally, X-ray is produced due to Compton scattering by the high-temperature plasma around the disk \citep{Kawashima2012}.
On the other hand, if the mass accretion rate is much lower than the Eddington rate, the spectral energy widely distributes in the range between radio and gamma rays \citep{Narayan1995, Manmoto1996, Yuan2003}.
Since the ionization and heating rate depend on the SED, the different SEDs could vary in flow structure and accretion rate.
Two-dimensional radiation hydrodynamics simulations on Bondi scale, in which the multi-temperature blackbody radiation is taken into consideration, showed that the critical gas density, above which a transition to super-Eddington accretion occurs, slightly decreases compared to the case of power-law spectra \citep{Takeo2018}. 
This is because photoionization for accretion disk spectra is less efficient than that for single power-law spectra. 
We need to study the Bondi-Hoyle-Lyttelton accretion considering more realistic SED.

In this study, we assume the cosine function for the angular distribution of the radiation field caused by accretion disks.
The disk could produce much stronger angular dependence if the mass accretion rate is above the Eddington rate \citep{Watarai2005, Ohsuga2005}.
Conversely, the angular dependence of the radiation field is weak if the mass accretion rate is much lower than the Eddington rate.
Also, the angular distribution varies with time as the accretion disk undergoes precession motion.
The multidimensional simulations of the accretion disk around the BHs are needed to obtain more realistic models of the radiation field \citep{Machida2000, Hawley2001, Ohsuga2009, Ohsuga2011}.
Two-dimensional radiation hydrodynamics simulations around a static BH have been performed with parameters such as the size of the shadow region and angular distribution of the anisotropic radiation field \citep[e.g.][]{Sugimura2017,Takeo2018}. 
They suggested the accretion rate becomes much higher in the case of a large shadow size and strong anisotropy. 
Investigation of the dependence of shadow size and anisotropy in the case of moving BHs is left as future work.

Outflow affects the gas flow around BHs.
A strong bipolar outflow has been observed around compact objects with mass accretion rates above the Eddington rate \citep[e.g.][]{Fabrika2004}.
Also, the jet has been detected in many sources \citep[e.g.][]{Walker2018}.
The previous studies showed that outflow pushes out the ambient gas, and accretion rate is reduced to approximately $20-30\%$ of the Bondi-Hoyle-Lyttleton accretion rate \citep[e.g.][]{Li2020,Bosch2022}.
They indicated that acceleration rate is $40-80\%$ smaller than without outflow.
For further study on the impacts of outflow, we will include these effects in future works.

\section{Summary and Conclusions}
\label{S:Summary and Conclusions}
In the present work, 
we study the Bondi-Hoyle-Lyttleton accretion mechanism onto IMBHs ($10^4M_{\odot}$) by three-dimensional radiation hydrodynamics simulations, taking into account the anisotropic radiation field originating from the accretion disk.
We consider the situation in which the gas with relatively low metallicity ($Z=0.1Z_{\odot}$) flows in perpendicular to the rotation axis of the accretion disk (parallel to the disk plane).
We take into account the radiation force acting on the dusty gas and decrease in the absorption coefficient due to the dust sublimation.
The luminosity of the accretion disk is supposed to increase with the mass accretion rate.
Our major findings are summarized as follows.

If the density of dusty gas is relatively high ($\sim 10^4~\rm cm^{-3}$) and the relative velocity between IMBHs and the gas is low ($\sim20~\rm km~s^{-1}$), time-averaged accretion rate is 0.6 \% of the Bondi-Hoyle-Lyttleton accretion rate
($4 \times 10^{-6}M_{\odot}\rm yr^{-1}$).
It is found that an ionized region like two spheres glued together at the equatorial plane appears around the IMBH and the dense shock shell is formed nearby the ionization front.
The radiation force around the rotation axis of the disk works to prevent gas accretion so that the gas mainly accretes through the disk equatorial plane.
The gravity of the dense shock shell accelerate the IMBHs at $\sim 10^{-8}\rm cm~s^{-2}$.
Note that the accretion rate periodically increases, but does not significantly affect the time-averaged accretion rate and acceleration.

In the case of high relative velocity ($\sim100~\rm km~s^{-1}$),
the accretion rate, $4 \times 10^{-6}M_{\odot}\rm yr^{-1}$, is approximately equal to the Bondi-Hoyle-Lyttleton accretion rate.
This is because the radiation force hardly prevents the gas accretion.
Even around the rotation axis of the disk, the radiation force is significantly small due to the dust sublimation.
The dense shock shell is not formed, and the IMBHs are decelerated due to the gas accretion.

When the gas density is extremely high ($\sim10^6~\rm cm^{-3}$), the time-averaged accretion rate is 0.6 \% of the Bondi-Hoyle-Lyttleton accretion rate, $\sim 4\times 10^{-4}M_{\odot}\rm yr^{-1}$. 
Due to the gravity by dense downstream gas, the IMBHs are decelerated although the dense shock shell is formed at the vicinity of the ionization front.
The accretion rate oscillates periodically as in the case of the density of $\sim 10^4~\rm cm^{-3}$. 
This is because when the surface density of the shell increases to a certain degree, part of the shell falls due to the gravity of the IMBHs.
Periodic variations in accretion rate occur even when the radiation is isotropic. This result differs from previous work, possibly because we have taken into account dust sublimation.

Based on the present results, the IMBHs are expected to continue floating in the early galaxies ($z \gtrsim 6$) at the velocity of $\gtrsim {\rm several} \times 10 \, {\rm km~s^{-1}}$ without significantly mass growth,
if the gas density of the galaxies is $\sim 10^4~\rm cm^{-3}$.
On the other hand, if the density of interstellar medium is extremely high, $\sim 10^6~\rm cm^{-3}$, 
even the IMBHs with the initial velocity of about 
$\sim 100 \, {\rm km~s^{-1}}$ will slow down due to momentum transport caused by the mass accretion.
If the Bondi accretion begins with decreasing the velocity, the IMBHs could grow rapidly.
This may lead to the evolution from IMBHs to SMBHs.

\section*{Acknowledgements}

We thank an anonymous referee for fruitful comments.
This work was supported by JSPS KAKENHI Grant Numbers 
JP22KJ0435 (E.O.), 
JP21H04488, JP18K03710(K.O.), 
JP20H04724, JP21H04489(H.Y.) and JP23K13139 (H.F.), and  Japan Society for the Promotion of Science and JST FOREST Program, Grant Number JPMJFR202Z (H.Y.). 
A part of this research has been funded by the MEXT as 
”Program for Promoting Researches on the Supercomputer Fugaku” 
(Toward an unified view of the universe: 
from large scale structure to planets, Grant Number JPMXP1020200109) 
(K.O., H.F., H.Y.), 
by MEXT as “Program for Promoting Researches on the Supercomputer Fugaku” (Structure and Evolution of the Universe Unraveled by Fusion of Simulation and AI; Grant Number JPMXP1020230406) and used computational resources of supercomputer Fugaku provided by the RIKEN Center for Computational Science (Project ID:
hp230204, K.O.),
and by Joint Institute for Computational Fundamental Science (JICFuS, K.O.).
Numerical computations were performed with computational resources 
provided by the Multidisciplinary Cooperative Research Program 
in the Center for Computational Sciences, University of Tsukuba, 
Oakforest-PACS operated by the Joint Center for 
Advanced High-Performance Computing (JCAHPC), 
Cray XC 50 at the Center for Computational Astrophysics (CfCA) of 
the National Astronomical Observatory of Japan (NAOJ), 
the FUJITSU Supercomputer PRIMEHPC FX1000 
and FUJITSU Server PRIMERGY GX2570 (Wisteria/BDEC-01) 
at the Information Technology Center, The University of Tokyo.

%%%%%%%%%%%%%%%%%%%%%%%%%%%%%%%%%%%%%%%%%%%%%%%%%%
\section*{Data Availability}
The data underlying this article will be shared on reasonable request to the corresponding author.



\bibliographystyle{mnras}
\bibliography{eo24_final}

%\begin{thebibliography}{99}
%\end{thebibliography}

%%%%%%%%%%%%%%%%%%%%%%%%%%%%%%%%%%%%%%%%%%%%%%%%%%

%%%%%%%%%%%%%%%%% APPENDICES %%%%%%%%%%%%%%%%%%%%%

\appendix
\section{Oscillation of accretion rate for isotropic model}

\begin{figure}
    \centering
	\includegraphics[width=\columnwidth]{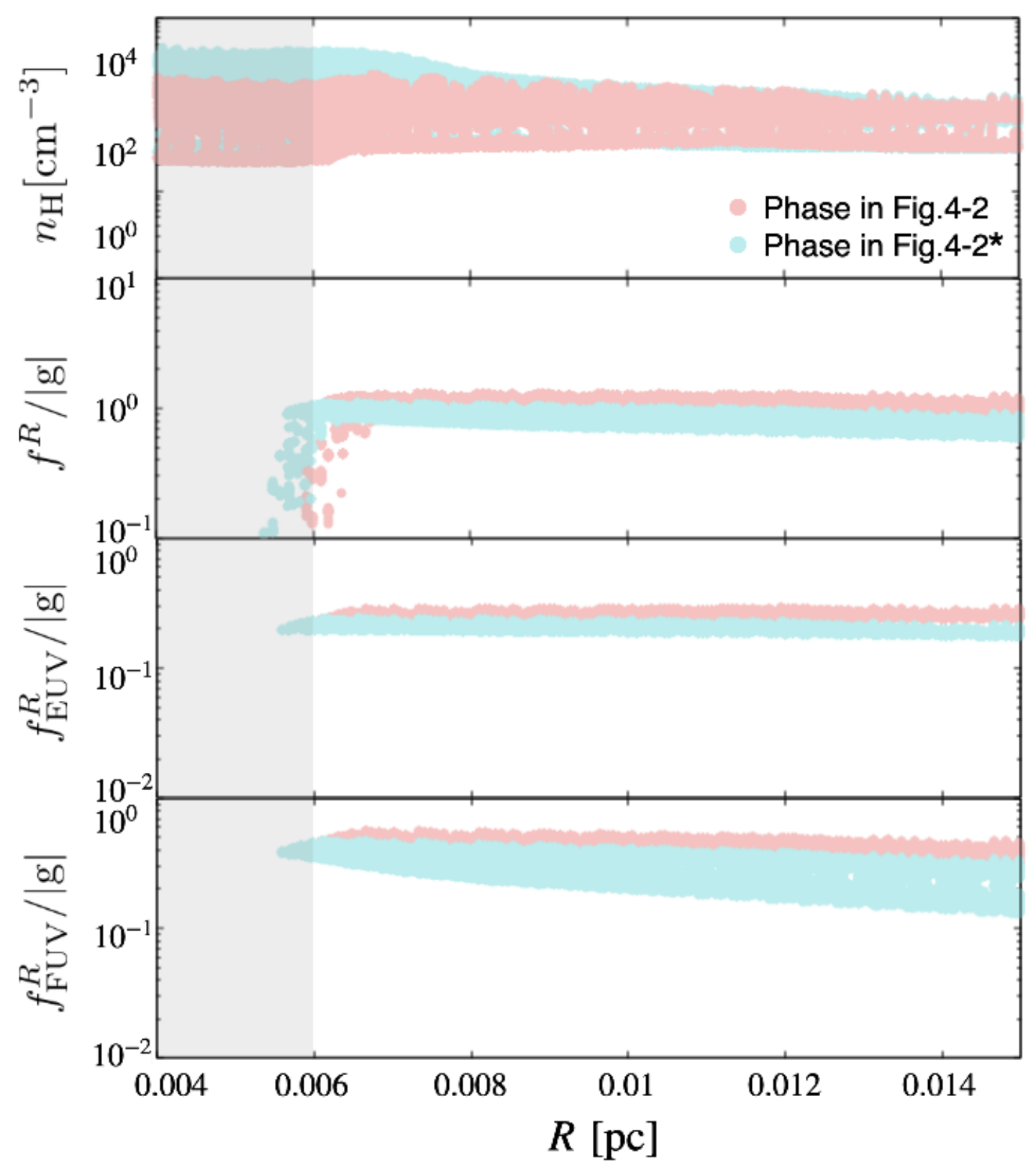}
    \caption{
    The number density $n_{\rm H}~\rm cm^{-3}$, radiation force $f^R, f_{\rm EUV}^R, f_{\rm FUV}^R$ normalized by gravity $\it{g}^R$ in Fig.\ref{fig:iso_peak}-2$^*$ (light-blue) and Fig.\ref{fig:iso_peak}-2 (pink).
    Here, $f^{R}$ and gray region means $f^{R}=f_{\rm EUV}^R+f_{\rm FUV}^R+f_{\rm IR}^R$ and region of dust sublimation, respectively.
    }
    \label{fig:iso_peak_appendix}
\end{figure}

Here we explain why the second of the two peaks appearing in the burst has a larger accretion rate (luminosity).
Figure \ref{fig:iso_peak} represents the number density ($n_{\rm H}$), ratio of radial component of total radiation force to gravity ($f^R/|g|$),
ratio of radial component of radiation force doe to EUV to gravity ($f^R_{\rm EUV}/|g|$),
and ratio of radial component of radiation force doe to FUV to gravity ($f^R_{\rm FUV}/|g|$) as a function of the distance from the IMBH ($R$).
The light-blue and pink mens the results at phase 4-2 ($t\sim 0.21715~\rm Myr$) and phase 4-2$^*$ ($t\sim 0.2174~\rm Myr$), respectively (see Fig. \ref{fig:iso_peak}).
Phase 4-2 roughly corresponds to the first peak, while phase 4-2$^*$ is a little before the second peak. The accretion rate in phase 4-2$^*$ is approximately equal to the accretion rate in Phase 4-2. 

It is found that the density is larger in the phase 4-2$^*$ than in phase 4-2 at slightly outside the dust sublimation region.
This difference in density produces a difference in radiation flux (radiation force) through attenuation by absorption.
Indeed, we find that the radiation force in the phase 4-2$^*$ is slightly weaker than that in the phase 4-2.
In phase 4-2, the accretion rate starts to decrease due to radiation force, but in phase 4-2$^*$, the accretion rate continues to increase further due to the weaker radiation force. 
Thus, the accretion rate (luminosity) is larger in second peak.
Note that the main component of radiation force is due to FUV.
The gas density is high in phase 4-2$^*$ because relatively large amount of gas that is not absorbed in the first peak remains around the BH. Conversely, the gas density in phase 4-2 is low because a large amount of gas is swallowed in the second peak.

%%%%%%%%%%%%%%%%%%%%%%%%%%%%%%%%%%%%%%%%%%%%%%%%%%

% Don't change these lines
\bsp	% typesetting comment
\label{lastpage}
\end{document}